\documentclass{PoS}
\usepackage{graphicx}
\usepackage{amsmath,amssymb,latexsym}
\usepackage{url}
\usepackage{bm}
\usepackage{xspace}
\usepackage{nicefrac}

\title{Origin of the ankle in the ultra-high energy
  cosmic ray spectrum and the extragalactic protons below it}

\ShortTitle{Origin of the ankle...}

\author{\speaker{Glennys R. Farrar}\\
 Center for Cosmology and Particle Physics,  New York University, NY 10003, USA \\
        E-mail: \email{gf25@nyu.edu}}

\author{Michael Unger\\
Center for Cosmology and Particle Physics,  New York University, NY 10003, USA\\
Institut f\"ur Kernphysik, Karlsruher Institut f\"ur Technologie,
        76021 Karlsruhe, Germany\\
E-mail: \email{mu495@nyu.edu}}

\author{Luis A. Anchordoqui\\
Department of Physics and Astronomy, Lehman College at CUNY, NY 10468,
USA \\
Department of Physics, Graduate Center, City University of New York, NY 10016, USA\\
Department of Astrophysics, American Museum of Natural History, NY 10024, USA\\
E-mail: \email{laa410@nyu.edu}}

\abstract{The sharp change in slope of the ultra-high energy cosmic
  ray (UHECR) spectrum around $10^{18.6}~{\rm eV}$ (the ankle),
  combined with evidence of a light but extragalactic component near
  and below the ankle which evolves to intermediate composition above,
  has proved exceedingly challenging to understand theoretically.  We
  show that for a range of source conditions, photo-disintegration of
  ultra-high energy nuclei in the region surrounding a UHECR
  accelerator naturally accounts for the observed spectrum and
  composition of the entire extragalactic component, which dominates
  above about $10^{17.5}~{\rm eV}$.  The mechanism has a clear
  signature in the spectrum and flavors of neutrinos.}

\FullConference{The 34th International Cosmic Ray Conference,\\
		30 July- 6 August, 2015\\
		The Hague, The Netherlands}

\def\EeV{\ifmmode {\mathrm{Ee\kern -0.07em V}}\else
                   \textrm{Ee\kern -0.07em V}\fi\xspace}
\def\GeV{\ifmmode {\mathrm{Ge\kern -0.07em V}}\else
                   \textrm{Ge\kern -0.07em V}\fi\xspace}
\def\TeV{\ifmmode {\mathrm{Te\kern -0.07em V}}\else
                   \textrm{Te\kern -0.07em V}\fi\xspace}
\def\eV{\ifmmode {\mathrm{e\kern -0.07em V}}\else
                   \textrm{e\kern -0.07em V}\fi\xspace}
\def\meV{\ifmmode {\mathrm{me\kern -0.07em V}}\else
                   \textrm{me\kern -0.07em V}\fi\xspace}
\newcommand{\energy}[1]{\ensuremath{10^{#1}}\,\eV}
\def\gcm{\ensuremath{\mathrm{g/cm}^2}\xspace}

\def\Sibyll{\textsc{Sibyll2.1}\xspace}
\def\Epos{\textsc{Epos-LHC}\xspace}
\def\QgII{\textsc{QGSJetII-04}\xspace}

\newcommand{\CRP} {{\scshape CRPropa}\xspace}
\newcommand{\bolddot}[1] { \overset{\bm .}{#1}}



\begin{document}

\section{Introduction}
The cosmic ray spectrum spans roughly eleven decades of energy,
$\energy{9} \lesssim E \lesssim \energy{20}$. Continuously running
monitoring using sophisticated equipment on high altitude balloons,
satellites and ingenious installations on the Earth's surface explore
the plummeting flux that decreases nearly three orders of magnitude
per energy decade until eventually suffering a strong suppression
around $10^{19.6}~{\rm eV}$~\cite{Abbasi:2007sv,Abraham:2008ru}. Close
examination reveals several features in the spectrum, the three most
important being: the steepening of the spectrum dubbed the ``knee''
occurring at approximately \energy{15.6}, from $J(E) \propto
E^{-2.7\pm 0.01}$ to $E^{-3.10 \pm 0.07}$ ~\cite{Antoni:2005wq}; a
less prominent ``second knee'', corresponding to a further softening
$J (E) \propto E^{-3.52 \pm 0.19}$ around
\energy{17.5}~\cite{AbuZayyad:2000ay}; a pronounced hardening of the
spectrum at $E \approx \energy{18.6}$, constituting the so-called
``ankle'' feature~\cite{Abbasi:2007sv,Abraham:2010mj}.

The variations of the spectral index reflect various aspects of cosmic
ray production, source distribution and propagation.  The first and
second knee have straightforward explanations, as reflecting the
maximum energy of Galactic magnetic confinement or acceleration
capability of the sources, both of which grow linearly in the charge
$Z$ of the nucleus; the first knee being where protons drop out and
the second knee where the highest-$Z$ Galactic CRs (GCRs) drop out.
As the energy increases above the 2nd knee to the ankle, the
composition evolves from heavy to light~\cite{Kampert:2012mx} while
the CR arrival directions are isotropic to high accuracy throughout
the range~\cite{Abreu:2011ve}.  Finally, as the energy increases above
the ankle, not only does the spectrum harden significantly, but the
composition gradually becomes heavier (interpreting the data using
conventional extrapolations of accelerator-constrained particle
physics models)~\cite{Aab:2014kda}.  The Telescope Array Collaboration
reports a composition consistent with pure proton, but their data are
also consistent with the results of the Pierre Auger Observatory when
systematic and statistical uncertainties are taken into account
\cite{Abbasi:2015xga}.

This observed evolution in the extragalactic CR (EGCR) composition and
spectral index pre\-sents a major conundrum.  A pure proton
composition might be compatible with the observed spectrum of
EGCRs~\cite{Berezinsky:2002nc} when allowance is made for experimental
uncertainties in the energy scale and the fact that the real local
source distribution is not homogeneous and
continuous~\cite{Ahlers:2012az} (although the sharpness of the ankle
is difficult to accommodate), but a pure proton composition is
incompatible with the depth-of-shower-maximum ($X_{\rm max}$)
distributions observed by Auger~\cite{Aab:2014kda} unless current
extrapolations of particle physics are incorrect.  On the other hand,
models which fit the spectrum and composition at highest energies,
predict a deep gap between the end of the GCRs and the onset of the
EGCRs.  Models can be devised to fill this gap, but fine-tuning is
required to position this new population so as to just fit and fill
the gap~\cite{Gaisser:2013bla}.

Here we offer a resolution to this conundrum, by showing that
``post-processing'' of UHECRs via photo-disintegration in the
environment surrounding the source, can naturally explain the entire
spectrum and composition.  In our model, EGCRs below the ankle are
predominantly protons knocked off higher energy nuclei in the region
surrounding the accelerator, and the spectrum and composition above
the ankle are predominantly dictated by the accelerator and
propagation to Earth.  The model makes distinctive predictions about
the spectrum and flavor ratios of neutrinos, which should enable it to
be tested.  If the ankle and the protons below it arise on account of our
mechanism, we obtain a new constraint on UHECR sources beyond the
Hillas criterion and total-energy-injection requirements, namely that
the environment around the source has the conditions giving rise to
the required amount of photo-disintegration.\\


\section {Formation of the Ankle}
\label{sec:ankle}
To illustrate the mechanism we have identified to create the ankle and
generate extragalactic protons at lower energies, consider a system in
which the accelerator is embedded in a photon field and the CRs are
trapped by magnetic fields in this environment.  It is standard to
consider the impact of the photons that are encountered during the
process of acceleration, in case those limit the maximum energy.  The
new feature here is to also allow for and explore the impact of
interactions with photons in the surrounding medium.  Examples could
be the dusty torus surrounding an AGN or the ISM of the star-forming
region surrounding most young pulsars.  See also
~\cite{Allard:2009fb}.

A simple analytic treatment is instructive.  Details are derived
in~\cite{Unger:2015laa}, but the essential simplifying assumptions are
i) a CR either escapes without changing energy, with a rate
$\tau_\mathrm{esc} $, or the CR interacts one or more times before
escaping; ii) no energy is lost except through an interaction and
whenever a nucleus interacts it loses one or more nucleons by
photodisintegration or photopion production; in this case the nucleus
loses a fraction of its energy corresponding to the reduction in its
nuclear mass; iii) $\tau_\mathrm{esc} $ and $\tau_\mathrm{int} $ are
independent of position and depend only on $E, \, A,\, Z$ of the
nucleus.

In this approximation the number of nuclei in a given energy bin with
specified $A,\, Z$ decreases exponentially with time. The decay
constant is $ \tau = (\nicefrac{1}{\tau_\mathrm{esc}} +
\nicefrac{1}{\tau_\mathrm{int}})^{-1}$, where $\tau_\mathrm{esc}$ and
$\tau_\mathrm{int}$ are the escape and interaction times respectively.
A fraction $\eta_\mathrm{esc}$ of the particles escape without
interaction
\begin{eqnarray}
   N_\mathrm{esc} & = & \int_{0}^\infty
   \frac{1}{\tau_\mathrm{esc}}\, N(t)\, dt
 =  \frac{N_0}{1 + \nicefrac{\tau_\mathrm{esc}}{\tau_\mathrm{int}}} \equiv N_0 \, \eta_\mathrm{esc}
\label{eq:nesc}
\end{eqnarray}
and the rest interact before escaping so $\eta_\mathrm{int} = 1 -
\eta_\mathrm{esc}$.  Note that $\eta_\mathrm{esc}$ and
$\eta_\mathrm{int}$ depend only on the ratio of the escape and
interaction times, but not on the absolute value of either of them.

The mechanism for generating an ankle-like feature can be explained
most easily by considering the full dissociation of injected nuclei of
mass $A^\prime$, to $A^\prime$ nucleons. If the escape and interaction
times are both power laws in energy, $\tau_\mathrm{esc} \equiv
a\,(E^\prime/E_0)^\delta$ and $\tau_\mathrm{int} \equiv
b\,(E^\prime/E_0)^\beta$, then
\begin{equation}
\eta_\mathrm{esc}(E^\prime) = \left(1 + R_0\,(E^\prime/E_0)^{\delta-\beta}\right)^{-1},
\label{eq:modi}
\end{equation}
where $R_0=a/b$ is the ratio of the escape and interaction time at
reference energy $E_0$. Therefore, if $\delta>\beta$ the propagation
in the source acts as a {\itshape low-pass filter} on the spectrum of
injected particles leading to a cutoff in the spectrum at high
energies. This situation typically occurs in stochastic shock
acceleration scenarios where $\delta>0$, for non-thermal photon energy
spectra with $\beta<0$. If, on the other hand, the escape times
decrease with energy, as in the case of diffusion in turbulent
magnetic fields, then it is possible to have $\delta<\beta$ leading to
a {\itshape high-pass filter} for the energy spectrum of injected
nuclei.  In this case, the lower the energy the more nuclei interact,
leading to an apparent hardening of the spectrum of escaping nuclei.
The produced nucleons on the other hand have energies of $E^* =
E^\prime / A^\prime$.  These nucleons are most abundant at low
energies and have a steeper spectrum $\varpropto (1 -
\eta_\mathrm{esc}(E^* \, A^\prime))$.  If interactions of these
nucleons are negligible, the high-pass scenario leads naturally to an
ankle-like feature separating the nucleonic fragments from the
remaining nuclei, at the energy where the number of escaping nuclei is
equal to the number of nucleons.

We consider two characteristic photon spectra. The
first consists of broken power-law as a
simplified representative of non-thermal emission.  The second is a
modified black-body spectrum.  The
interaction time for a nucleus with Lorentz factor
propagating through the photon field with differential number density
$n(\epsilon)$ is given by~\cite{Stecker:1969fw}.   Inspired by
the energy dependence of the diffusion coefficient for the propagation
in a turbulent magnetic field, we model $\tau_\mathrm{esc}$ as a power
law in rigidity $E/Z$

\begin{equation}
   \tau_\mathrm{esc} = \tau_0 \left(\frac{E/Z}{E_0}\right)^\delta.
\end{equation}

\begin{figure}[t!]
    \includegraphics[clip, viewport = 8 9 180 122, width=0.49\linewidth]{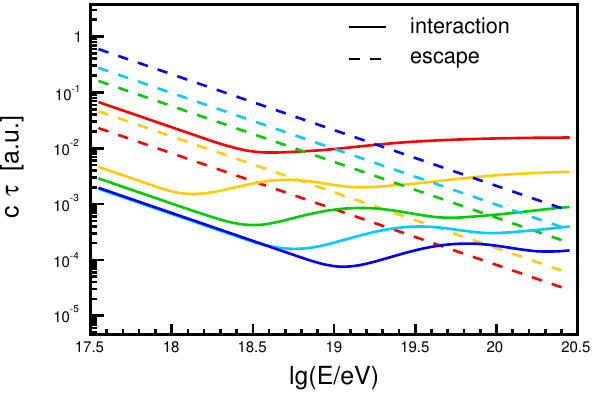}\quad
  \includegraphics[clip, viewport = 197 145 369 260, width=0.49\linewidth]{Fig1.pdf}
\caption[source]{Left: Interaction and escape times for $A^\prime = 1,\, 4,\, 14,\, 26$ and $56$ (bottom to top for escape and vice versa for interaction).
Injected $^{28}$Si flux (bold dashed) and escaping flux.
Right: Mass groups are color-coded with increasing mass from left to
right. The thin black solid line denotes the sum of
all escaping nuclei. Nucleons from photo-dissociation are shown as a
thin dashed curve and nucleons from photo-pion-production as dotted
curve.}
\label{fig:esc}
\end{figure}


The left panel of Fig~\ref{fig:esc} shows the escape and interaction times as a
function of the CR energy (proton, He, N, Si and Fe). We adopt $\delta
= -1$, as for Bohm diffusion. We take a broken
power-law photon spectrum, which peaks in the far infrared
($\varepsilon_0 = 50$~\meV) and has slope parameters $\alpha =
+\frac{5}{2}$ and $\beta = -2$. The normalization of the photon
density and escape time, $n_0$ and $\tau_0$, are chosen so that the
ratio of interaction and escape time at \energy{19} for iron nuclei,
$R_{19}^{\rm Fe}$, is 275; the interaction times are calculated
including both photo-disintegration and photo-pion production.

The gross features of the energy dependence of the interaction times
can be understood in the approximation of resonant interactions at
center of mass energies of $\varepsilon^{\mathrm{CM}}_\mathrm{res}$.
At low cosmic-ray energies, reaching
$\varepsilon^{\mathrm{CM}}_\mathrm{res}$ requires high photon energy
($\varepsilon > \varepsilon_0$), so that the interaction time
decreases with cosmic-ray energy as $\tau \varpropto E^{\beta +1}$.
However, for high enough CR energy, the resonance can be reached in
collisions with photons of $\varepsilon < \varepsilon_0$; from here,
as the CR energy increases, the photon density decreases as
$\varepsilon^{\alpha}$, and correspondingly the interaction times
increase. The laboratory energy of the inflection point of the
interaction times for a CR nucleus of mass $M$ is at $E = M
\varepsilon_\mathrm{res} / (2 \varepsilon_0)$. The inflection point of
the photo-dissociation times can be seen as a dip in the right panel
of Fig.~\ref{fig:esc}, e.g.\ at around \energy{19.3} for iron
nuclei. At slightly higher energy photo-pion production becomes
important, with the result that the energy dependence of the
interaction time is roughly speaking an L-shaped curve in a log-log
presentation.

Using these energy-dependent interaction and escape times,
we propagate nuclei through the source environment
with the procedure described in~\cite{Unger:2015laa}.

As an example, we injected silicon nuclei, $A^\prime=28$,
with a power law $E^{-1}$ and maximum energy
$Z^\prime E^p_{\rm max} = 3\times 10^{19}$~\eV. The unmodified spectrum
and the spectrum of escaping
nuclei are shown Fig.~\ref{fig:esc}; the units are arbitrary, but the normalization in the two figures is the
same. At low energies, the injected nuclei are depleted because
$\tau_{\rm esc} \gg \tau_{\rm int}$, but the escaping nuclei follow
the original spectral index, because in this example the interaction
and escape times are parallel, as to be expected for $\delta = \beta +
1$. Once the corner of the L-shape is reached, the fraction of
escaping nuclei grows leading to an apparent hardening of the spectral
index.

Even for the simple case of an injection of a single nuclear species
into the source environment, we obtain a complex evolution of the
mass composition with energy. At low energies the composition is
dominated by knock-off nucleons whereas at high energies the
composition becomes heavier as the ratio of escape to interaction time
drops and more heavy nuclei can escape before interacting.

A qualitatively similar result is obtained when assuming a modified
black-body spectrum, where the parameters $T=150$~K and $\sigma=2$ have been chosen to match the peak
and shape of the power-law spectrum used above.  The corresponding
interaction lengths and escaping spectra are shown in~\cite{Unger:2015laa}.  In
both cases the spectrum of escaping particles displays an ankle-like
feature from the transition between the soft spectrum of nucleons and
the hard spectrum of nuclei.  We have checked that the approximation of neglecting single-nucleon energy loss in the source environment is valid for the examples and parameter ranges discussed here.  For a concrete astrophysical realization of this scenario, one must also check and if necessary include the effect of interactions with UHECRs with hadrons in the source environment.\\

 \begin{figure*}[t!]
   \includegraphics[width=\linewidth]{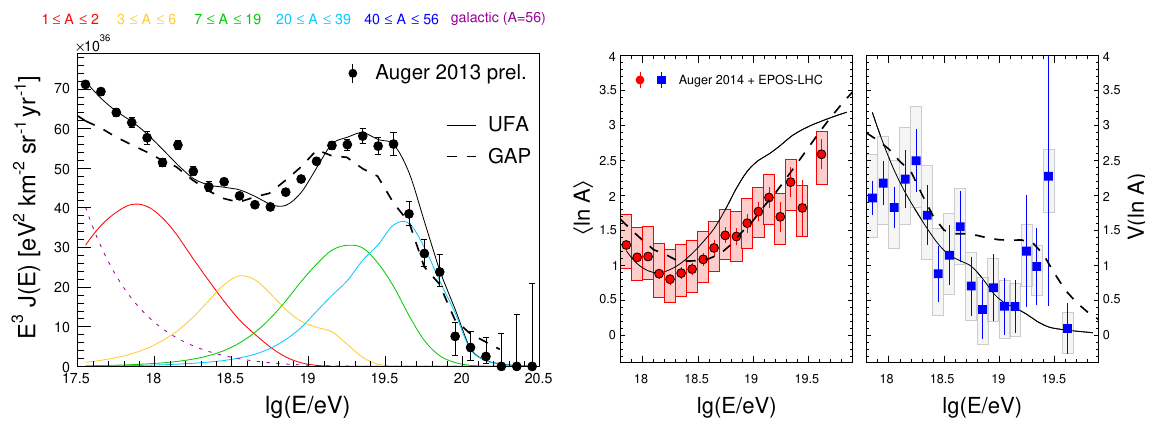}
 \caption[earth]{Spectrum and composition at Earth. Dots are data from the
 Pierre Auger Observatory~\cite{ThePierreAuger:2013eja, Aab:2014kda},
 error bars denote the statistical uncertainties and the shaded boxes
 in the center and right figure illustrate the experimental systematic uncertainties
 of the moments of $\ln A$. These composition estimates are based on an
 interpretation of air shower data with \Epos.  The solid lines denote the
 predictions of our model and the dashed lines are the results from~\cite{Globus:2015xga}.}
 \label{fig:fit}
 \end{figure*}

\section{Comparison to Data}
The model presented here is a very simple one, yet even so it offers a remarkably good accounting for the flux and composition at Earth as determined by the Pierre Auger
Observatory.   We compare to the measured flux from \energy{17.5} to above
\energy{20}~\cite{ThePierreAuger:2013eja} and the mean
and variance of the distribution of the logarithm of mass on top of
the atmosphere, $\langle \ln A \rangle$ and $V(\ln
A)$~\cite{Aab:2014kda, Abreu:2013env}.

We take as fiducial example, a hard UHECR injection spectrum $E^{-1}$
as expected for acceleration in young neutron
stars~\cite{Blasi:2000xm} and use the broken power law photon spectrum
with peak $\epsilon_{0}$ in the far infrared at $\varepsilon_0 =
50$~\meV and spectral indices $\alpha = +\frac{5}{2}$ and $\beta =
-2$. We assume that the evolution of the sources follows the star
formation rate~\cite{Robertson:2015uda} (see \cite{Taylor:2015rla} for
a model of cosmic rays above the ankle using a negative source evolution).

To keep the complexity of the model to a minimum here, we inject only a single nuclear
species.  The best description of the data is obtained with
$^{28}$Si.  Another free parameter is the maximum energy, for which the best
fit gives a value of $Z^\prime \times \energy{18.5} = 3\times 10^{19}$~\eV.

The ratio of the escape time and the interaction time in the
photo-disintegration region at a specified rigidity and composition is
the only free parameter characterizing the photo-disintegration; the
best fit corresponds to iron at $E=10^{19}$~eV having an escape length
275 times its interaction length.  The spectral index and
normalization of the Galactic spectrum are free ``nuisance''
parameters as well. The best fit gives a spectral index of -4.2 and a
contribution of about half of the flux at the lowest energy of the fit
(\energy{17.5}), fully compatible with indications from
KASCADE-Grande~\cite{Apel:2013ura}.

The resulting fit is shown in comparison to data in
Fig.~\ref{fig:fit}. There is a good overall agreement between the
model and the data. The shape of the spectrum is described well,
including the ankle and the flux suppression. The model also
qualitatively reproduces the increase of the average logarithmic mass
with energy and the decrease of its variance.  Normalizing this model
to the observed flux at Earth, we infer a comoving energy injection
rate in CRs at $z=0$ and above \energy{17.5} of
$\bolddot{\epsilon}_{17.5} = 8.2 \times 10^{44}$ $\frac{\rm erg}{\rm
Mpc^{3}\, yr}$.

To explore the sensitivity of the quality of the fit to uncertainties
in the model and data, we have varied the parameters of the model and
the general features are consistently reproduced (see
\cite{Unger:2015laa} for details). Changing the extragalactic
background light model from~\cite{Kneiske:2003tx} which represents an
upper limit, to the lower limit of \cite{Kneiske:2010pt}, results in a
similar description of the data with $A^\prime = 23$ and an
$R_{19}^{\rm Fe}$ increased by factor of $1.5$. Fixing $A^\prime = 28$
and letting the spectral index float freely leads to a spectrum
following $E^{-1.04}$.  Changing the photon field from a broken power
  law to a modified black body prefers the mass to be at $A^\prime =
  27$ and reduces $R_{19}^{\rm Fe}$ by 0.7.  Concerning the photon
  field, we checked that we get acceptable fits for a peak photon
  energy in the range of 25 to 100 m\eV, values of $\beta$ in range of
  -2.5 to -1.5, and that there is little sensitivity on $\alpha$.  For
  the escape lengths, the power law index $\delta$ can be as large as
  $-0.7$. Even slower escapes could probably be accommodated by a
  variation of $\beta$ at the same time.

We find that the best fit is obtained within the experimental
systematics when shifting the energy scale up by $+1\,\sigma_{\rm
  sys.}  = 15\%$ and by shifting $\langle \ln A \rangle$ and $V(\ln
A)$ corresponding to a shift of the shower maximum by $-1\,\sigma_{\rm
  sys.} \approx -10~\gcm$.

Using alternative models for the hadronic interactions of air showers
(\Sibyll or \QgII instead of \Epos), decreases the value of the
$\langle \ln A \rangle$ data points by about $\langle \ln A \rangle =
-0.6$ and leads to a worse fit of the data, within this limited
exploration of parameter space.  It is worthwhile noting that if the
difference between models gives a fair estimate of the uncertainties
of the mass determination in both directions,
$\sigma_\mathrm{theo}(\langle \ln A \rangle) = \pm 0.6$, then a
hadronic interaction model that would give a larger mass estimate than
\Epos would make this fiducial model fit even better.

A related analysis was presented in \cite{Globus:2015xga}.
They explore a
specific model of GRB internal shock acceleration and find
that for sufficiently strong source evolution and properly chosen GRB
model parameters, photo-disintegration on photons created within the
GRB, can produce an ankle-like feature. The predictions of this
model are shown as dahsed lines in Fig.~\ref{fig:fit}.

The neutrino signals of our model are discussed
in~\cite{Unger:2015laa}.  We emphasize the distinctive $\bar{\nu}_{e}$
enrichment due to beta decay of spallated neutrons.

\section{Conclusions}
We have exposed
general conditions which give rise to an ankle-like feature in the
source emission spectrum, with light composition below and
increasingly heavy composition above, as called for by the data with
conventional particle physics modeling.  We illustrated the high
quality of the fit which can be obtained to the data, with a fiducial
model in which nuclei accelerated up to a maximum rigidity, with
spectrum $\propto E^{-1}$, are subject to photo-disintegration in the
vicinity of the accelerator before escaping for their journey to
Earth.  Such a scenario can be reasonably achieved in astrophysical
sources, as will be discussed in a future publication.  Very
generally, when nuclei remain trapped in the turbulent magnetic field
of the source environment their escape time can decrease faster with
increasing energy than does their interaction time.  When these
general conditions are realized, only the highest energy particles can
escape the source environment unscathed.  In other words, the source
environment acts as a high-pass filter on the spectrum of cosmic rays.
Nuclei below the crossover energy scatter off the far-infrared photons
in the source environment, with ejection of nucleons or alpha
particles.  This produces a steep spectrum of secondary nucleons.  The
superposition of the steeply falling nucleon spectrum with the harder
spectrum of the surviving nuclear fragments creates an ankle-like
feature in the total source emission spectrum.  This occurs at an
energy of about \energy{18.6}, for a large range of parameter values,
reflecting the kinematics of the giant dipole resonance.  The spectrum
above the ankle exhibits a progressive transition to heavy nuclei, as
the escape of non-interacting nuclei becomes efficient.  Abundant
production of $\bar{\nu}_{e}$'s is a signature of this mechanism.

\section*{Acknowledgments}
We would like to acknowledge many useful discussions with our
colleagues of the Pierre Auger Collaboration. Furthermore we thank
David Walz for his support regarding questions about \CRP.  MU
acknowledges the financial support from the EU-funded Marie Curie
Outgoing Fellowship, Grant PIOF-GA-2013-624803. The research of GRF is
supported in part by the U.S. National Science Foundation (NSF), Grant
PHY-1212538 and the James Simons Foundation; she thanks KIPAC/SLAC for
their hospitality. The research of LAA is supported by NSF (Grant
CAREER PHY-1053663) and NASA (Grant NNX13AH52G); he thanks the Center
for Cosmology and Particle Physics at New York University for its
hospitality.

\end{document}